\documentclass{styles/svproc}
\usepackage{graphicx}
\usepackage{multicol}
\usepackage{footmisc}
\usepackage[utf8]{inputenc}             % para los acentos
\usepackage[caption=false]{subfig}      % para las subfiguras. El caption=false es solo para sacar el warning (igual anda sin eso)
\usepackage{graphicx}

\usepackage{url}

\let\svthefootnote\thefootnote
\newcommand\freefootnote[1]{%
	\let\thefootnote\relax%
	\footnotetext{#1}%
	\let\thefootnote\svthefootnote%
}

\begin{document}
\mainmatter              % start of a contribution
\title{Checkpoint and Restart: An Energy Consumption Characterization in Clusters}
\titlerunning{Checkpoint and Restart}  % abbreviated title (for running head)
%                                     also used for the TOC unless
%                                     \toctitle is used
%
\author{Marina Morán\inst{1} \and Javier Balladini\inst{1} \and
Dolores Rexachs\inst{2} \and Emilio Luque\inst{2}}
\authorrunning{Marina Morán et al.} % abbreviated author list (for running head)
%
%%%% list of authors for the TOC (use if author list has to be modified)
\tocauthor{Marina Moran, Javier Balladini, Dolores Rexachs and Emilio Luque}
\institute{Facultad de Informática, Universidad Nacional del Comahue, Buenos Aires 1400, 8300 Neuquén, Argentina\\
\email{marina@fi.uncoma.edu.ar},
\and
Departamento de Arquitectura de Computadores y Sistemas Operativos, Universitat Autònoma de Barcelona, Campus UAB, Edifici Q, 08193 Bellaterra (Barcelona), España}

\maketitle              % typeset the title of the contribution

\begin{abstract}
The fault tolerance method currently used in High Performance Computing (HPC) is the rollback-recovery method by using checkpoints. This, like any other fault tolerance method, adds an additional energy consumption to that of the execution of the application. The objective of this work is to determine the factors that affect the energy consumption of the computing nodes on homogeneous cluster, when performing checkpoint and restart operations, on SPMD (Single Program Multiple Data) applications. We have focused on the energetic study of compute nodes, contemplating different configurations of hardware and software parameters. We studied the effect of performance states (states P) and power states (states C) of processors, application problem size, checkpoint software (DMTCP) and distributed file system (NFS) configuration. The results analysis allowed to identify opportunities to reduce the energy consumption of checkpoint and restart operations.
% We would like to encourage you to list your keywords within
% the abstract section using the \keywords{...} command.
\end{abstract}
\section{Introduction}

\freefootnote{The final authenticated version is available online at https://doi.org/10.1007/978-3-030-20787-8\_2}
High Performance Computing (HPC) continues to increase its computing power while increasing its energy consumption. Given the limitations that exist to supply energy to this type of computers, it is necessary to know the behavior of energy consumption in these systems, to find ways to limit and decrease it. In particular, for the exaescala era, a maximum limit of 20 MW is estimated \cite{Bergman08exascalecomputing}.

The fault tolerance method most currently used in HPC is the rollback-recovery method by using checkpoints. This, like any other fault tolerance method, adds an additional energy consumption to the execution of the application \cite{Shalf2010}. Due to this, it is important to know and predict the energy behavior of fault tolerance methods, in order to manage their impact on the total energy consumption during the execution of an application.

The objective of this work, which is an extension of \cite{Moran2018}, is to determine the factors that affect the energy consumption of checkpoint and restart operations (C/R), on Single Program Multiple Data (SPMD) applications on homogeneous cluster, contemplating different configurations of hardware and software parameters. A cluster system has compute nodes, storage nodes, and at least one interconnection network. We have focused on the energetic study of computing nodes, and we have extended the previous work contemplating, on the one hand, a second experimentation platform, and on the other hand, the storage node, in particular when studying the impact of checkpoint files compression. The energy consumption of the network is not considered in this article.

The contributions of this article are:
\begin{itemize}
\item A study of the system's own factors (hardware and software) and of applications factors that impact the energy consumption produced by checkpoint and restart operations.

\item The identification of opportunities to reduce the energy consumption of checkpoint and restart operations.
\end{itemize}

Section \ref{sec:Trabajos-Relacionados} presents some related works, while in section \ref{sec:Identificacion-de-Factores} the factors that affect the energy consumption are identified. Section \ref{sec:plataforma} mention the experimental platform and the design of experiments, whose results and their analysis are presented in the section \ref{sec:experimentos}. Finally, the conclusions and future work can be found in the section \ref{sec:conclusiones}.%

\section{Related work}\label{sec:Trabajos-Relacionados}
\cite{Diouri} and \cite{Meneses} evaluates the energetic behavior of the coordinated and uncoordinated C/R with message logs. In \cite{Meneses} they also evaluate the parallel recovery and propose an analytic model to predict the behavior of these protocols at exascale. In \cite{Mills2014} they use an analytic model to compare the execution time and the consumed energy of the replication and the coordinated C/R. \cite{Amrizal2017} and \cite{Dauwe2017} presents an analytic model to estimate the optimal interval of a multilevel checkpoint in terms of energy consumption. They do not measure dissipated power but use values from other publications. In \cite{Diouri2013} they measure the power dissipated and the execution time of the high level operations involved in checkpoint (coordinated, uncoordinated and hierarchical) varying the number of cores involved. They do not use different processor frequencies, nor do they indicate whether the checkpoint is compressed or not. In \cite{Rajachandrasekar2015}, \cite{Cui2016} and \cite{Saito2013} a framework for energy saving of the C/R are presented. In \cite{Rajachandrasekar2015}, many small I/O operations are replaced by a few large, single-core operations to make the checkpoint and restart more energy efficient. They use RAPL to measure and limit energy consumption. \cite{Cui2016} propose to have a core to execute a replica of all the processes of the node in order to avoid re-execution from the last checkpoint and analytically compare the energy consumption of this proposal with the traditional checkpoint. In \cite{Saito2013} a runtime that allows modifying the clock frequency and the number of processes that carry out the I/O operations of the C/R are designed, in order to optimize the energy consumption. Another work that analyze the impact of the dynamic scaling of frequency and voltage on the energy consumption of checkpoint operations is in \cite{Mills2013}. They measure the power at the component level while writing the checkpoint locally and remotely and compare variations in remote storage: NFS using the kernel network stack and NFS using the IB RDMA interface. \cite{ibtesham2014} evaluate the energy consumption of an application that uses compressed checkpoints. They show that when using compression, more energy is spent but time is saved, so that the complete execution of the application with all its checkpoints can be benefited from an energy point of view.

Our work focuses on coordinated C/R at the system level. The dissipated power of the checkpoint and restart operations are measurements obtained with an external physical meter. We have not found papers that evaluate the impact of C states and NFS configurations on the energy consumption of C/R operations.

\section{Factors that affect the energy consumption}\label{sec:Identificacion-de-Factores}
Energy can be calculated as the product between power and time. Any factor that may affect one of these two parameters should be considered and then analyze how it affects energy consumption. These factors belong to different levels: Hardware, Application Software and System Software.

\subsection{Hardware}\label{subsec:hardware}
The Advanced Configuration and Power Interface (ACPI) specification provides an open standard that allows the operating system to manage the power of the devices and the entire computing system \cite{acpi}. It allows managing the energy behavior of the processor, the component that consumes the most energy in a computer system \cite{Silveira2016}. ACPI defines Processor Power States (Cx states), where C0 is the execution state, and C1...Cx are inactive states.

A processor that is in the C0 state will also be in a Performance State (Px states). The status P0 means an execution at the maximum capacity of performance and power demand. As the number of state P increases, its performance and demanded power is reduced. Processors implement P states using the Dynamic Frequency and Voltage Scaling technique (DVFS) \cite{le2010dynamic}. Reducing the voltage supplied reduces the energy consumption. However, the delay of the logic gates is increased, so it is necessary to reduce the clock frequency of the CPU so that the circuit works correctly. In certain multicore processors, each core is allowed to be in a different P state.

When there are no instructions to execute, the processor can be put in a C state greater than 0 to save energy. There are different C state levels, where each of the levels could turn off certain clocks, reduce certain voltages supplied to idle components, turn off the cache memory, etc. The higher the C state number, the lower the power demanded, but the higher the latency required to return to state C0 (execution status). Some processors allow the choice of a C state per core.

As both states, C and P, have an impact on power and time, it is necessary to evaluate their impact on energy consumption during C/R operations.

The GNU/Linux kernel supports frequency scaling through the subsystem CPUFreq (CPU Frequency scaling). This subsystem includes the scaling governors and the scaling drivers. The different scaling governors represent different policies for the P states. The available scaling governors are: performance (this causes the highest frequency defined by the policy), power save (this causes the lowest frequency defined by the policy), userspace (the user defines frequency), schedutil (this uses CPU utilization data available from the CPU scheduler), ondemand (this uses CPU load as a CPU frequency selection metric) and conservative (same as ondemand but it avoids changing the frequency significantly over short time intervals which may not be suitable for systems with limited power supply capacity). The scaling drivers provide information to the scaling governors about the available P states and make the changes in those states \footnote{https://www.kernel.org/doc/html/v4.14/admin-guide/pm/cpufreq .html}. In this work we use userspace and ondemand governors.

\subsection{Application software}\label{subsec:sw-aplicacion}
Basically, a checkpoint consists in saving the state of an application, so that in case of failure it can restart the execution from that saved point. The larger the problem size of the application, the longer the time required to save its state. Its incidence, at least over time, converts problem size into a factor that affects the energy consumption of C/R operations.

\subsection{System software}\label{subsec:sw-sistema}
There are two types of system software highly involved in C/R operations. On the one hand, the system that carry out these operations. On the other hand, since the checkpoint file needs to be protected in stable and remote storage, it is necessary to use a distributed file system. In our case, the system software we use is Distributed MultiThreaded CheckPointing (DMTCP) \cite{DMTCP} and Network File System (NFS). Both have configuration options that affect the execution time and/or power, and therefore are factors that affect the energy consumption of the C/R operations.

In the NFS case, folders can be mounted synchronously (sync option) or asynchronously (async option). If an NFS folder is mounted with the sync option, writes at that mount point will cause the data to be completely downloaded to the NFS server, and written to persistent storage before returning control to the client\footnote{https://linux.die.net/man/5/nfs}. Thus, the time of a write operation is affected by varying this configuration.

In the case of DMTCP, it is a tool that performs checkpoints transparently on a group of processes spread among many nodes and connected by sockets, as is the case with MPI (Message Passing Interface) applications. DMTCP is able to compress (using the gzip program) the state of a process to require less disk storage space and reduce the amount of data transmitted over the network (between the compute node and the storage node). The use or not of compression impacts the time and the power required, therefore it is another factor that affects energy consumption.

\section{Experimental platforms and design}\label{sec:plataforma}

\subsection{Experimental platform}
The experiments were carried out on two platforms. Platform 1, on
which most of the analysis of this work is carried out, is a cluster of computers with a 1 Gbps Ethernet network. Each node, both computing and storage, has 4 GiB of main memory, a SATA hard disk of 500 GB and 7200 rpm, and an Intel Core i5-750 processor, with a frequency range of 1.2 GHz to 2.66 GHz (with the Intel Turbo Boost \footnote{https://www.intel.com/content/www/us/en/architecture-and-technology/turbo-boost/turbo-boost-technology.html} disabled), four cores (without multithreading), 8 MiB of cache and 95 W TDP. The clock frequency range goes from 1.199 GHz. to 2.667 GHz. Platform 2 is a compute node connected to a storage node with a 1 Gbps Ethernet network. The compute node has an Intel Xeon E5-2630 processor, a frequency range of 1.2 GHz to 2.801 GHz (with the Intel
Turbo Boost mechanism disabled), six cores (with multithreading disabled),
16 GiB of main memory, 15 MiB of cache and TDP of 95 W. It uses a
Debian 9 Stretch operating system. The storage computer has an Intel Core 2 Quad Q6600 processor, four cores (without multithreading), 8 GiB of main memory and 4 MiB of cache memory. The clock frequency range goes from 1.2 GHz. to 2.801 GHz.

The nodes of Platform 1 and the computing node of Platform 2 use the
GNU/Linux operating system Debian 8.2 Jessie (kernel version 3.16 of 64 bits), OpenMPI version 1.10.1 as an MPI message passing library, and the tool checkpoint DMTCP version 2.4.2, configured to compress the checkpoint files. The network file system used to make the remote writing of the checkpoint files is NFS v4 (Network File System).

For power measurements we use the PicoScope 2203 oscilloscope (whose
accuracy is 3\%), the TA041 active differential probe, and the PP264
60 A AC/DC current clamp, all Pico Technology products. The electrical
signals captured by the two-channel oscilloscope are transmitted in
real time to a computer through a USB connection. The voltage is measured
using the TA041 probe that is connected to an input channel of the
oscilloscope. The current of the phase conductor, which provides energy
to the complete node (including the power source) is measured using
the current clamp PP264, which is connected to the other input channel
of the oscilloscope.

The selected application\footnote{https://computing.llnl.gov/tutorials/parallel\_comp/\#ExamplesHeat} for system characterization is a SPMD heat transfer application written in MPI that uses the float data type. This application describes, by means of an equation, the change of temperature in time over a plane, given an initial temperature distribution and certain edge conditions.

\subsection{Experimental design}
Two compute nodes are used (unless otherwise specified) and each compute node writes to a dedicated storage node through an NFS configured in asynchronous mode (unless otherwise specified). The sampling rate used for both channels
of the oscilloscope was set at 1000 Hz. The power measurements correspond
to the power dissipated by the complete node including the source.
The tests were performed with the processor C states option active (unless otherwise specified).
For the measurements of the checkpoint and restart time we use the
option provided by DMTCP. To change the frequency of the processor,
the userspace governor is used. The same frequency is used in all cores at the same time. 

Each experiment consists of launching the application with one process
per core, letting it run during a preheating period (20 seconds),
performing a checkpoint manually, aborting the application from the
DMTCP coordinator and re-starting the application from the command
line with the script generated by DMTCP. The experiment is repeated
three times for each frequency and problem size due to the low
variability of the measurement instruments used.

\section{Experiments and results analysis}\label{sec:experimentos}
In this section we show the experiments and analyze how the mentioned factors influence power, time, and energy consumption of CR operations.

The first subsections (\ref{subsec:realpower} to \ref{subsec:problemSize}) analyze the effect of clock frequency (see subsection \ref{subsec:hardware}) and problem size (see subsection \ref{subsec:sw-aplicacion}) on Platforms 1 and 2. As there are no significant differences between the two platforms for these factors, we have carried out the study of the remaining factors, NFS configuration and compression of the checkpoint files (subsections \ref{subsec:nfs} and \ref{subsec:compresion}) only on Platform 1. We understand that these last factors are influenced mainly by the network, which is the same in both platforms.

\subsection{Real power}\label{subsec:realpower}

\begin{figure} 
    \centering
  \subfloat[Platform 1 at 2.667 GHz.]{%
       \includegraphics[width=0.47\linewidth]{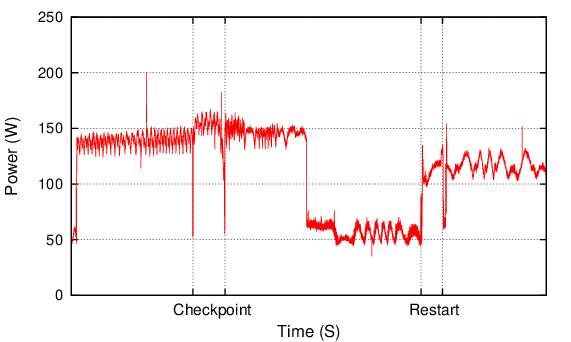}}
    \hfill
  \subfloat[Platform 2 at 2.801 GHz.]{%
        \includegraphics[width=0.47\linewidth]{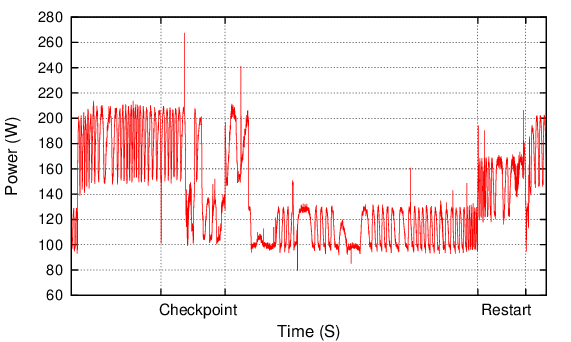}}
    \\
  \caption{Power Dissipation during Checkpoint and Restart.
  \label{fig:potencia_real} }
\end{figure}

In order to know the energy we need to know the average power dissipated of each operation. Fig. \ref{fig:potencia_real} shows the real power on both Platforms, obtained with the measurements delivered by the oscilloscope, when an application is executed, a checkpoint is done, a fault is injected, and a restart is initiated. These power measurements are averaged over the duration of the checkpoint and restart operation. In the rest of the work, when referring to dissipated power, we refer to the average dissipated power.

\begin{figure} 
    \centering
  \subfloat[Platform 1 at 2.667 GHz.]{%
       \includegraphics[width=0.47\linewidth]{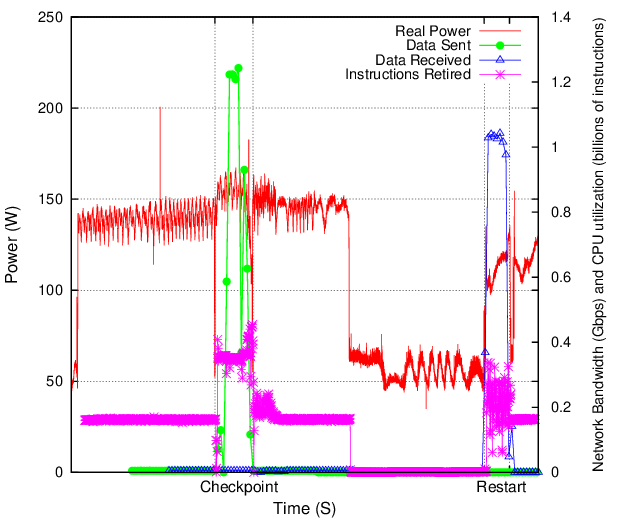}}
    \hfill
  \subfloat[Platform 2 at 2.801 GHz.]{%
        \includegraphics[width=0.47\linewidth]{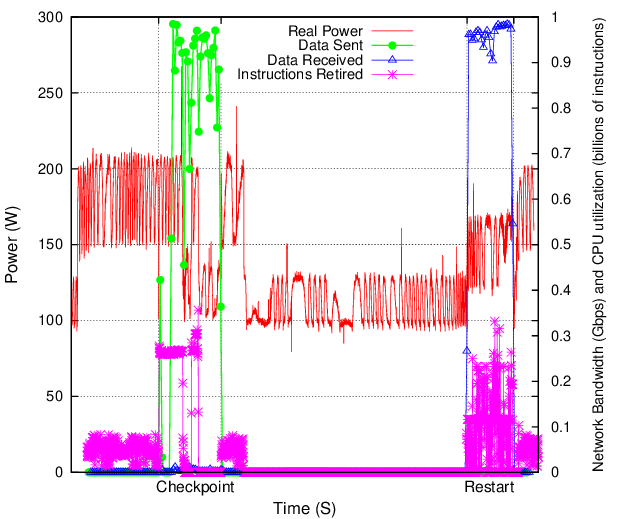}}
    \\
  \caption{Power Dissipation, Network Bandwidth and CPU utilization during Checkpoint and Restart.
  \label{fig:potencia_real_cpu_red} }
\end{figure}

Platform 2, unlike Platform 1, has two high phases and two low phases during the checkpoint. To know what happened in those phases we measured the CPU and network usage in both platforms, which is shown in Fig. \ref{fig:potencia_real_cpu_red}. When observing the use of CPU, we see that the high power phases coincide with a higher CPU usage, and that low phases coincide with a low CPU usage. During a checkpoint, the CPU is used to compress. From this it follows that in Platform 2, DMTCP, at times, stops compressing. When observing the use of the network, we see that on both platforms the transmission begins a few seconds after the checkpoint has started and continues until the end of the checkpoint. We also note that in the first low phase of Platform 2, the transmission rate drops by half, causing inefficient use of the network. Studying these inefficiencies could be part of future work. In the case of the restart, the transmission rate remains stable throughout its duration, on both platforms.

\subsection{Processor's P states} \label{subsec:estadosP}
\begin{figure} 
    \centering
  \subfloat[Dissipated Power]{%
       \includegraphics[width=0.47\linewidth]{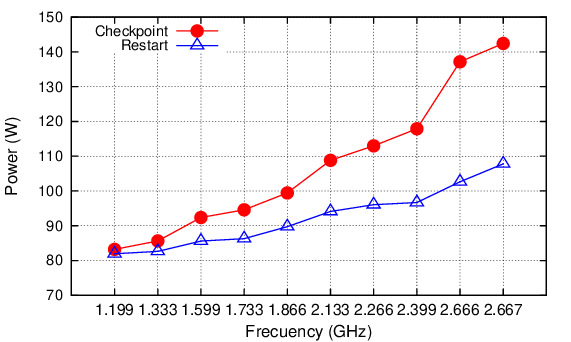}}
    \hfill
  \subfloat[Time]{%
        \includegraphics[width=0.47\linewidth]{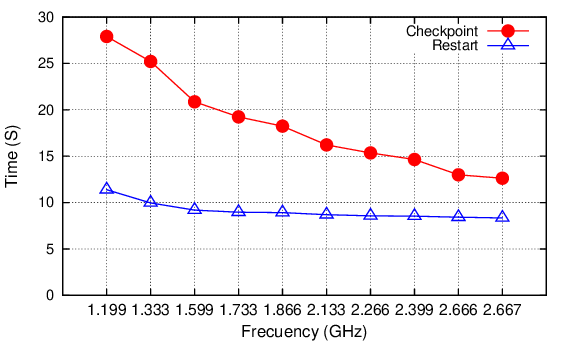}}
    \\
  \caption{Influence of P states on Platform 1.}
  \label{fig:incidencia_estados_P_plat1} 
\end{figure}

\begin{figure} 
    \centering
  \subfloat[Dissipated Power]{%
       \includegraphics[width=0.47\linewidth]{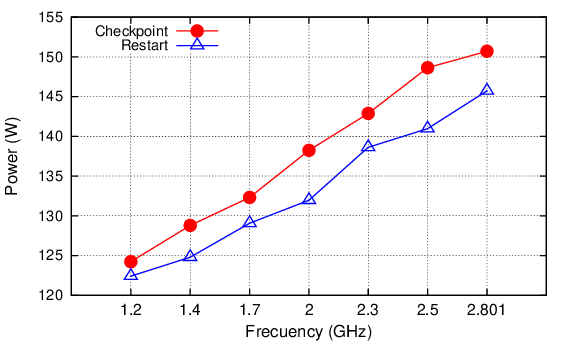}}
    \hfill
  \subfloat[Time]{%
        \includegraphics[width=0.47\linewidth]{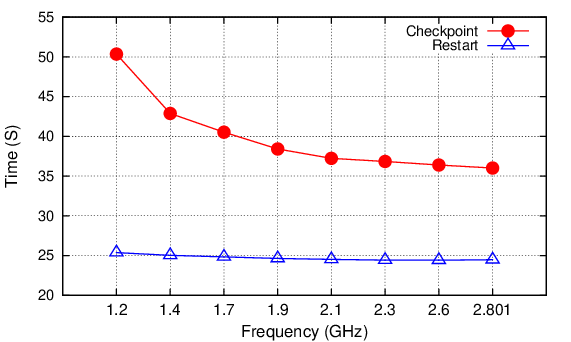}}
    \\
  \caption{Influence of P states on Platform 2.}
  \label{fig:incidencia_estados_P_plat2} 
\end{figure}

The impact of processor's P states on the C/R operations energy consumption was evaluated. The figures \ref{fig:incidencia_estados_P_plat1} and \ref{fig:incidencia_estados_P_plat2} show the average dissipated power (a) and time (b) for different frequencies of the processor, on both platforms. It is observed that as the clock frequency increases, the dissipated power increases and the time decreases. The functions obtained are strictly increasing for the case of  dissipated power and strictly decreasing for the case of time. It is also observed that the checkpoint is more affected by clock frequency changes, both in the power dissipated and in time.

\subsection{Processor's C states}\label{subsec:estadosC}

\begin{figure} 
    \centering
  \subfloat[Checkpoint]{%
       \includegraphics[width=0.47\linewidth]{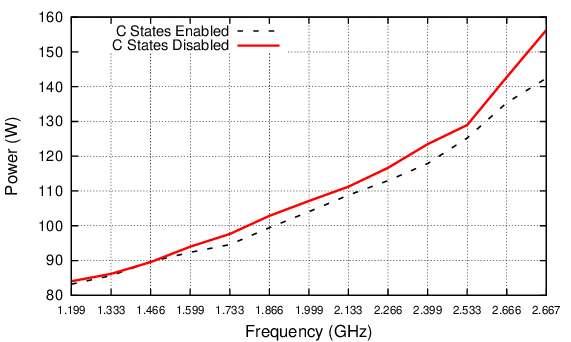}}
    \hfill
  \subfloat[Restart]{%
        \includegraphics[width=0.47\linewidth]{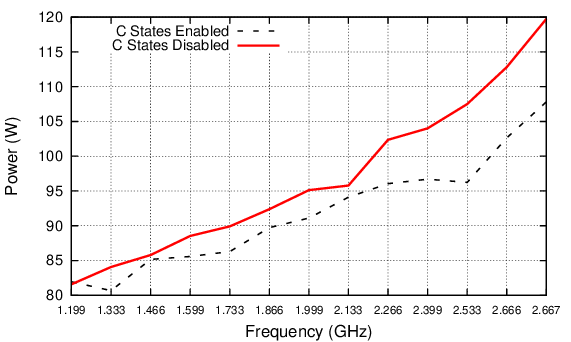}}
    \\
  \caption{Influence of C states on power dissipation - Platform 1.}
  \label{fig:incidencia_estados_C_plat1} 
\end{figure}

\begin{figure} 
    \centering
  \subfloat[Checkpoint]{%
       \includegraphics[width=0.47\linewidth]{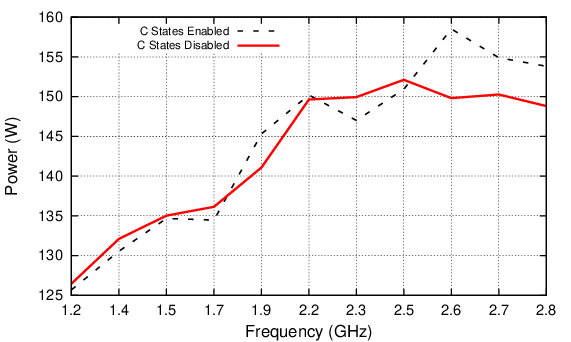}}
    \hfill
  \subfloat[Restart]{%
        \includegraphics[width=0.47\linewidth]{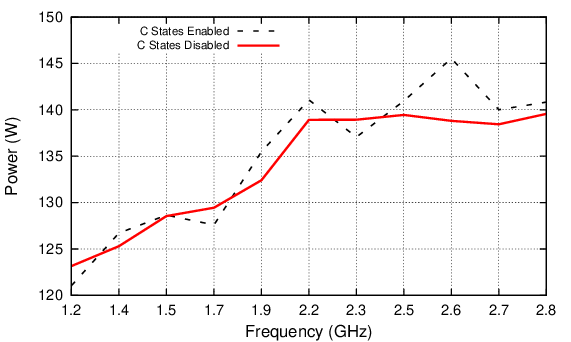}}
    \\
  \caption{Influence of C states on power dissipation - Platform 2.}
  \label{fig:incidencia_estados_C_plat2} 
\end{figure}

During the writing or reading of a checkpoint file it is possible that the processor has idle moments and therefore transitions between C states can occur. These transitions can affect the power dissipation of the processor. To study its behavior, C/R operations were performed with the C states enabled and disabled, for several frequencies of the processor, on Platform 1 and 2. The results are shown in the figures \ref{fig:incidencia_estados_C_plat1} and \ref{fig:incidencia_estados_C_plat2}.

In both platforms it is observed that the power measurements with the C states enabled show greater variability, especially in the restart at Platform 1. It is also observed how the difference between the power dissipated with C states enabled and disabled increases with the increasing frequency of the processor. On Platform 1, this difference becomes approximately 9 \% for the checkpoint (at the maximum frequency), and 10\% for the restart (at the frequency 2.533 GHz). In Platform 2 these differences are smaller, 6\% for the checkpoint, and 5\% for the restart, in the frequency 2.6 GHz in both cases.

The consumption of energy is mainly benefited by the use of C states. On Platform 1, for the checkpoint case, the consumption is up to 13\% higher when the C states are disabled, and in the restart case, up to 20\% higher. On Platform 2 these differences are smaller, up to 5\%, except for the 1.4 GHz frequency, where the differences are greater than 15\%, both for checkpoint and restart. 

In any case, the best option is to keep the C states enabled
since they reduce the energy consumption by up to 13\% for the checkpoint,
and up to 20\% for the restart, at certain CPU frequencies. Times showed no variation
when enabling or disabling the C states.

\subsection{Problem size}\label{subsec:problemSize}

The impact of the problem size on energy consumption of C/R operations was evaluated. The power dissipated and the time of the checkpoint and restart were measured, for different problem sizes, on both platforms. 

\begin{figure} 
    \centering
  \subfloat[Power]{%
       \includegraphics[width=0.47\linewidth]{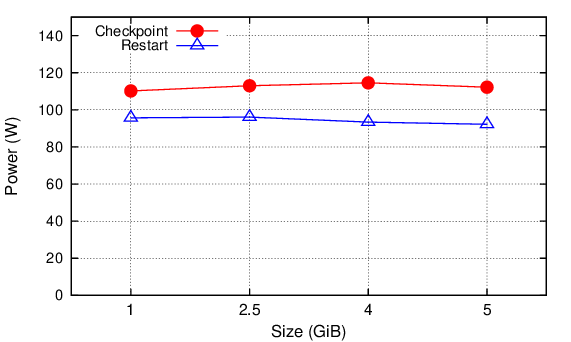}}
    \hfill
  \subfloat[Time]{%
        \includegraphics[width=0.47\linewidth]{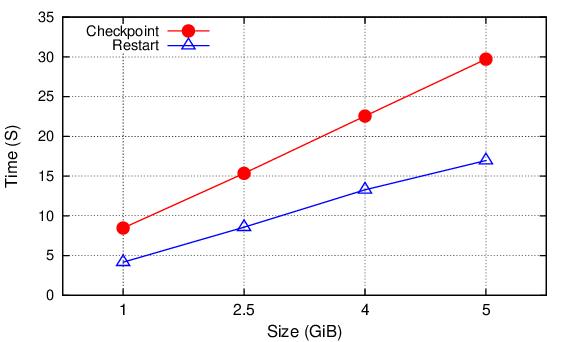}}
    \\
  \caption{Influence of problem size on Platform 1}
  \label{fig:size_plat1} 
\end{figure}

\begin{figure} 
    \centering
  \subfloat[Power]{%
       \includegraphics[width=0.47\linewidth]{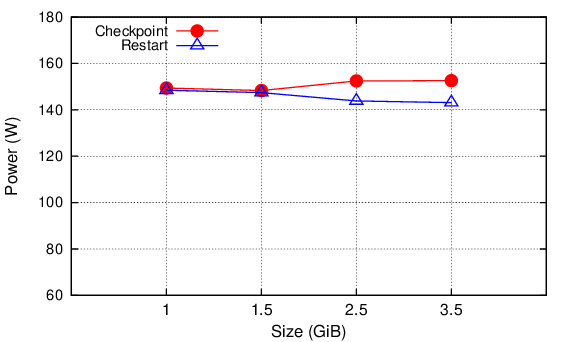}}
    \hfill
  \subfloat[Time]{%
        \includegraphics[width=0.47\linewidth]{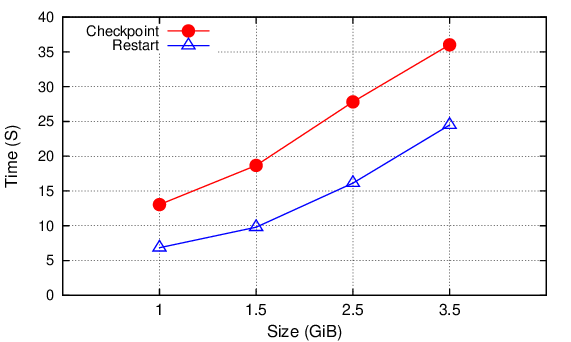}}
    \\
  \caption{Influence of problem size on Platform 2}
  \label{fig:size_plat2} 
\end{figure}

Problem sizes do not exceed the main memory available in the compute node. In the figures \ref{fig:size_plat1}(a) and \ref{fig:size_plat2}(a) it is observed that the power dissipated almost does not vary when varying the problem size (the differences do not exceed 4\% in any case). In the figures \ref{fig:size_plat1}(b) and \ref{fig:size_plat2}(b) it is observed that the time increases as the problem size increases, as expected.

\subsection{NFS configuration}\label{subsec:nfs}

\begin{figure} 
    \centering
  \subfloat[Power]{%
       \includegraphics[width=0.33\linewidth]{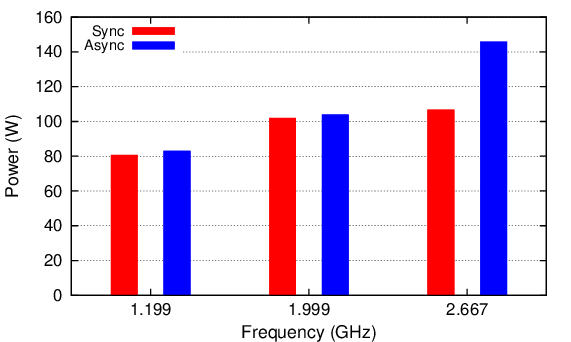}}
    \hfill
    \subfloat[Time]{%
        \includegraphics[width=0.33\linewidth]{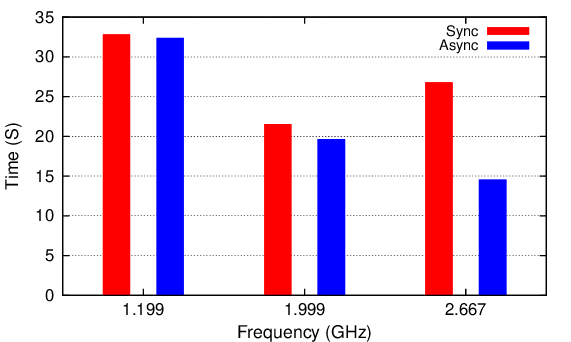}}
\subfloat[Energy]{%
        \includegraphics[width=0.33\linewidth]{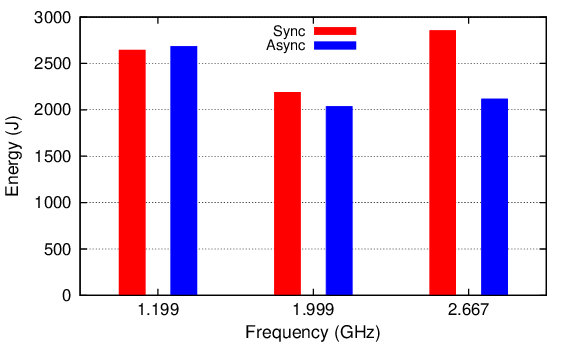}}
 \caption{Influence of sync/async configuration on power, time and energy consumption during checkpoint}
  \label{fig:sync_async} 
\end{figure}

The impact on energy consumption of the use of the NFS sync and async options was evaluated. The figure \ref{fig:sync_async} compares the dissipated power, time and energy consumed by a checkpoint  stored on a network file system mounted with the option sync and async, for three different clock frequencies (minimum, average and maximum of the processor), on Platform 1. For the three frequencies, the dissipated power is greater and the execution time is shorter when the asynchronous configuration is used. The shortest time is explained by the operating mode of the asynchronous mode, which does not need to wait for the data to be downloaded to the server to advance. The greater dissipated power may be due to the fact that the asynchronous mode decreases idle times of the processor, and therefore C states of energy saving does not activate (see subsection \ref{subsec:hardware}). However, for the minimum and average frequency, these differences are small, resulting in a similar energy consumption, as shown in Fig. \ref{fig:sync_async}(c). The asynchronous mode consumes 1.5\% more energy at the minimum frequency. It could be said that this configuration of the NFS does not affect the energy consumption when using this frequency. In the medium frequency, the asynchronous mode consumes 7\% less energy.
If we now observe what happens at the maximum frequency, we see that the asynchronous mode consumes 25\% less energy. Although the dissipated power is 37\% higher, the time is 85\% lower, and this means that at the maximum frequency, it is convenient to use the asynchronous mode for lower energy consumption.

\subsection{Checkpoint file compression}\label{subsec:compresion}

The impact of checkpoint files compression on the energy consumption of the computation node was analyzed, and in this work, we added the study of the storage node. The experiments were performed on a single computation node writing on a storage node, for three different clock frequencies (minimum, average and maximum available in the processor) and for the governor ondemand (see section \ref{subsec:hardware}), on Platform 1. 

\begin{figure} 
    \centering
  \subfloat[Power]{%
       \includegraphics[width=0.47\linewidth]{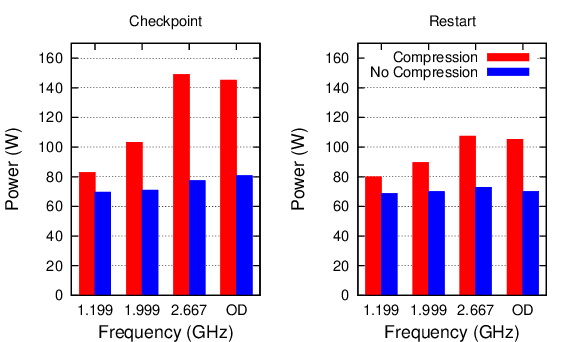}}
    \hfill
  \subfloat[Time]{%
        \includegraphics[width=0.47\linewidth]{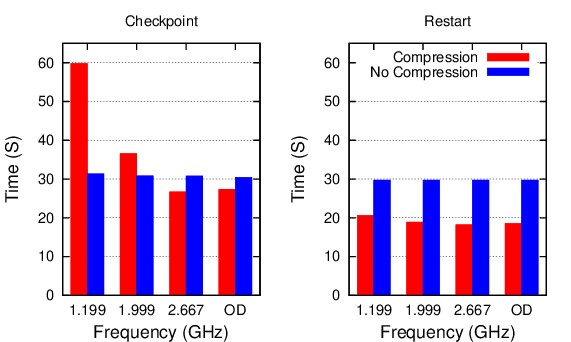}}
    \\
  \caption{Influence of compression on power and time.}
  \label{fig:compresionPyT} 
\end{figure}

The figure \ref{fig:compresionPyT} shows the power and time of checkpoint and restart, with and without compression of the checkpoint files.

The results obtained show the following:

\begin{itemize}
    \item The dissipated power, both in checkpoint and restart, is greater when using compression. This is due to the higher CPU usage that is required to run the compression program (gzip) that DMTCP uses. 
    \item Without compression, checkpoint and restart times almost do not vary for different clock frequencies. 
\end{itemize}

\begin{figure} 
    \centering
  \subfloat[Compute Node]{%
       \includegraphics[width=0.49\linewidth]{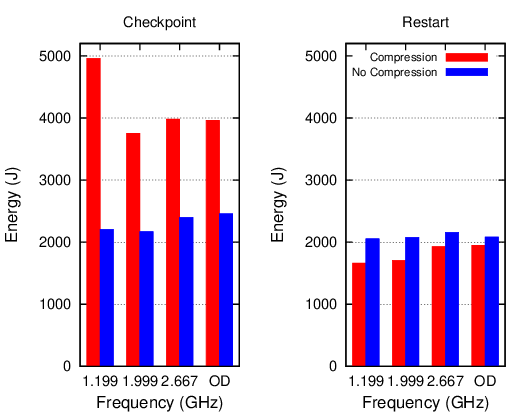}}
    \hfill
  \subfloat[Storage Node]{%
        \includegraphics[width=0.49\linewidth]{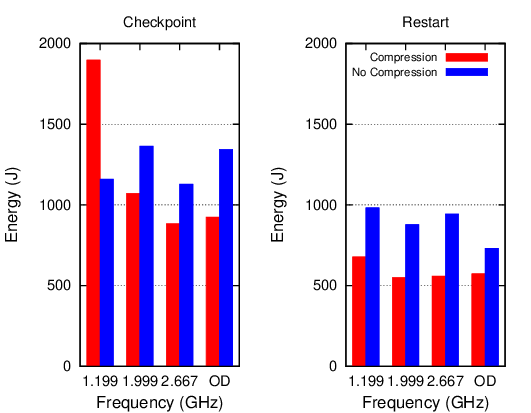}}
    \\
  \caption{Influence of compression on energy consumption.}
  \label{fig:compresionComputoAlmac} 
\end{figure}

Fig. \ref{fig:compresionComputoAlmac} shows the energy consumption of computation and storage nodes. In the case of the storage node, the clock frequencies indicated are the frequencies used by the computing node. Because the application does not share the storage node with other applications, the energy considered for the storage node is calculated using the total dissipated power (base power plus dynamic power). In the computation node, the energy consumption of the checkpoint is always greater when using compression (up to 55\% higher in the minimum frequency). In the storage node, the energy consumption of the checkpoint is always lower when using compression, except for the minimum frequency. The energy consumption of the restart is always lower when using compression (up to 20\%), both in computing and storage node.

\begin{figure}
  \includegraphics[width=\linewidth]{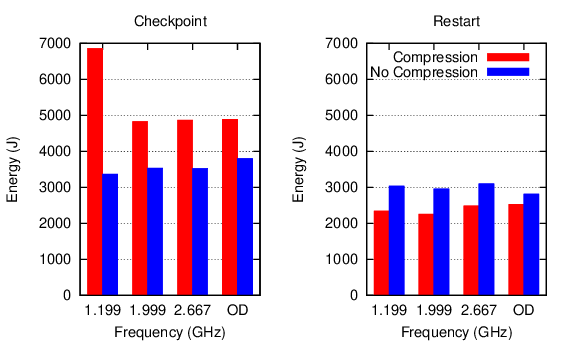}
  \caption{Total energy consumption: Compute node and storage node}
  \label{fig:compresionEnergiaTotal}
\end{figure}

Let's see the total energy consumption, considering both nodes (computation and storage), in Fig. \ref{fig:compresionEnergiaTotal}. For the checkpoint case, it is never advisable to compress, and even less at the minimum frequency. Although, when no compression is used, the minimum frequency is the most convenient, because it is the one with the lowest energy consumption. In any case, the energy consumption when no compression is used is similar for all frequencies studied, with differences that do not exceed 12\%.
For the restart case, it is always convenient to compress. When using compression, the lowest energy consumption is obtained with the frequency 1.999 GHz. However, by not using compression, the lowest energy consumption is obtained with the ondemand governor. Studying this behavior can be part of future work.

Taking into account that, in general, more checkpoint operations than restart operations are carried out, it is advisable not to use compression to reduce energy consumption.

\section{Conclusions and future work}\label{sec:conclusiones}

This work shows, from a series of experiments on an homogeneous cluster executing an SPMD application, how different factors of the system and the applications impact in the energy consumption of the C/R coordinated operations using DMTCP. The impact of the P and C states of the processor was studied. It was found that checkpoint operations are more sensitive to changes in P states than restart operations. On the contrary, the changes of the C states of the processor affect more the restart. In the evaluated platforms, the use of C states allows energy savings (up to 15\% for the checkpoint and 20\% for the restart). The increase in the application problem size under study always results in an increase in energy consumption, due to its high incidence in the time taken by the operations. It was found that by using the maximum clock frequency, up to 25\% energy savings was possible when using the asynchronous mode in the NFS configuration. The compression of the checkpoint files is beneficial for the restart. When considering only the computation node, up to 20\% of energy saving was registered when using compressed files. However, compression negatively impacts the checkpoint operation, with a 55\% higher energy consumption when using the minimum clock frequency. Among the future works it is expected to evaluate other factors that may affect the energy consumption of checkpoint and restart operations, such as the compression program used for checkpoint files, the parallel application programming model and the fault tolerance tool used.

%
% ---- Bibliography ----
%
%\bibliographystyle{plain}
%\bibliography{citas.bib}

@inproceedings{Ibtesham2014,
  title={Coarse-grained energy modeling of rollback/recovery mechanisms},
  author={D. Ibtesham, D. DeBonis, D. Arnold and K. B. Ferreira},
  booktitle={Dependable Systems and Networks (DSN), 2014 44th Annual IEEE/IFIP International Conference on},
  pages={708--713},
  year={2014},
  organization={IEEE}
}
@inproceedings{ibtesham2014coarse,
  title={Coarse-grained energy modeling of rollback/recovery mechanisms},
  author={Ibtesham, Dewan and DeBonis, David and Arnold, Dorian and Ferreira, Kurt B},
  booktitle={Dependable Systems and Networks (DSN), 2014 44th Annual IEEE/IFIP International Conference on},
  pages={708--713},
  year={2014},
  organization={IEEE}
}

@inproceedings{Saito2013,
  title={Energy-aware I/O optimization for checkpoint and restart on a NAND flash memory system},
  author={Saito, T., Sato, K., Sato, H., Matsuoka S.},
  booktitle={Proceedings of the 3rd Workshop on Fault-tolerance for HPC at extreme scale},
  pages={41--48},
  year={2013},
  organization={ACM}
}


@INPROCEEDINGS{Cui2016,
author = {X. Cui, T. Znati and R. Melhem},
booktitle = {2016 Intl IEEE Conferences on Ubiquitous Intelligence \& Computing, Advanced and Trusted Computing, Scalable Computing and Communications, Cloud and Big Data Computing, Internet of People, and Smart World Congress(UIC-ATC-ScalCom-CBDCom-IoP-SmartWorld)},
title = {Adaptive and Power-Aware Resilience for Extreme-Scale Computing},
year = {2016},
volume = {00},
number = {},
pages = {671-679},
keywords={Shadow mapping;Checkpointing;Fault tolerance;Fault tolerant systems;Resilience;Hardware},
doi = {10.1109/UIC-ATC-ScalCom-CBDCom-IoP-SmartWorld.2016.0111},
url = {doi.ieeecomputersociety.org/10.1109/UIC-ATC-ScalCom-CBDCom-IoP-SmartWorld.2016.0111},
ISSN = {},
month={July}
}
@inproceedings{Shalf2010,
  title={Exascale computing technology challenges},
  author={J. Shalf, S. Dosanjh and J. Morrison},
  booktitle={International Conference on High Performance Computing for Computational Science},
  pages={1--25},
  year={2010},
  organization={Springer}
}

@inproceedings{Amrizal2017,
  title={Optimizing Energy Consumption on HPC Systems with a Multi-Level Checkpointing Mechanism},
  author={M. A. Amrizal and H. Takizawa},
  booktitle={Networking, Architecture, and Storage (NAS), 2017 International Conference on},
  pages={1--9},
  year={2017},
  organization={IEEE}
}

@inproceedings{Dauwe2017,
  title={Optimizing checkpoint intervals for reduced energy use in exascale systems},
  author={D. Dauwe, R. Jhaveri, S. Pasricha, A. Maciejewski, and H.J. Siegel},
  booktitle={2017 Eighth International Green and Sustainable Computing Conference (IGSC)},
  pages={1--8},
  year={2017},
  organization={IEEE}
}
@article{Young1974,
    author = {J. W. Young},
    title = {{A first order approximation to the optimum checkpoint interval}}, 
    journal = {Comm. of the ACM, vol. 17, no. 9,1974},
    year = {1974},
}
@article{daly2006higher,
  title={A higher order estimate of the optimum checkpoint interval for restart dumps},
  author={J. T. Daly},
  journal={Future generation computer systems},
  volume={22},
  number={3},
  pages={303--312},
  year={2006},
  publisher={Elsevier}
}
@inproceedings{Miyoshi2002,
  title={Critical power slope: understanding the runtime effects of frequency scaling},
  author={A. Miyoshi, C. Lefurgy, E. Van Hensbergen, R. Rajamony and R. Rajkumar},
  booktitle={Proceedings of the 16th international conference on Supercomputing},
  pages={35--44},
  year={2002},
  organization={ACM}
}
@article{El-Sayed2015,
  title={Understanding practical tradeoffs in HPC checkpoint-scheduling policies},
  author={N. El-Sayed and B. Schroeder},
  journal={IEEE Transactions on Dependable and Secure Computing},
  volume={15},
  number={2},
  pages={336--350},
  year={2018},
  publisher={IEEE}
}
@article{El-Sayed2014,
    author = {N. El-Sayed and B. Schroeder},
    title = {{To checkpoint or not to checkpoint: Understanding energy-performance-i/o tradeoffs in hpc checkpointing}},
    journal = {Cluster Computing (CLUSTER), 2014 IEEE International Conference on},
    year = {2014},
}

@article{Tang2016,
    author = {K. Tang, D. Tiwari, S. Gupta, P. Huang, Q. Lu, C. Engelmann and X. He},
    title = {{Power-capping Aware Checkpointing: On the Interplay among Power-capping, Temperature, Reliability, Performance, and Energy}},
    journal = {IEEE/IFIP International Conference on Dependable Systems and Networks},
    year = {2016},
}
@article{Meneses2014,
  title={Energy profile of rollback-recovery strategies in high performance computing},
  author={E. Meneses, O. Sarood, and L. V. Kal{\'e}},
  journal={Parallel Computing},
  volume={40},
  number={9},
  pages={536--547},
  year={2014},
  publisher={Elsevier}
}

@article{Rajachandrasekar2015,
  title={Power-Check: An Energy-Efficient Checkpointing Framework for HPC Clusters},
  author={R. Rajachandrasekar, A. Venkatesh, K. Hamidouche and D. K. Panda},
  journal={2015 15th IEEE/ACM International Symposium on Cluster, Cloud and Grid Computing},
  year={2015},
  pages={261-270}
}

@INPROCEEDINGS{Diouri2013,
author = {M. Diouri, O. Gluck, L. Lefevre and F. Cappello},
booktitle = {2013 13th IEEE/ACM International Symposium on Cluster, Cloud and Grid Computing (CCGrid)(CCGRID)},
title = {ECOFIT: A Framework to Estimate Energy Consumption of Fault Tolerance Protocols for HPC Applications},
year = {2013},
volume = {00},
number = {},
pages = {522-529},
keywords={Protocols;Energy consumption;Fault tolerance;Fault tolerant systems;Checkpointing;Power demand;Calibration},
doi = {10.1109/CCGrid.2013.80},
url = {doi.ieeecomputersociety.org/10.1109/CCGrid.2013.80},
ISSN = {},
month={May}
}

@article{Aupy2013,
    author = {G. Aupy, A. Benoit, T. Herault, Y. Robert and J. Dongarra},
    title = {{Optimal Checkpointing Period: Time vs. Energy}},
    journal = {Performance Modeling, Benchmarking and Simulation of High Performance Computer Systems, Nov 2013, Denver, United States. <hal-00926199>},
    year = {2013},
}
@inproceedings{Mills2014,
  title={Energy consumption of resilience mechanisms in large scale systems},
  author={B. Mills, T. Znati, R. Melhem, K. B. Ferreira and R. E. Grant},
  booktitle={Parallel, Distributed and Network-Based Processing (PDP), 2014 22nd Euromicro International Conference on},
  pages={528--535},
  year={2014},
  organization={IEEE}
}

@article{Balladini_Muresano2013,
    author = {J. Balladini, R. Muresano, R. Suppi, D. Rexachs and E. Luque},
    title = {{Methodology for predicting the energy consumption of SPMD application on virtualized environments}},
    journal = {Computer Science \& Technology},
  issue_date = {December 2013},
   year = {2013},
 volume = {13},
 number = {1},
 month = dec,
 issn = {1666-6046},
 pages = {130--136},
 publisher = {Universidad Nacional de La Plata, Facultad de Informática}
}
@inproceedings{Meneses,
 author = {E. Meneses, O. Sarood, and L. V. Kale},
 title = {Assessing Energy Efficiency of Fault Tolerance Protocols for HPC Systems},
 booktitle = {Proceedings of the 2012 IEEE 24th International Symposium on Computer Architecture and High Performance Computing},
 series = {SBAC-PAD '12},
 year = {2012},
 isbn = {978-0-7695-4907-1},
 pages = {35--42},
 numpages = {8},
 acmid = {2419757},
 publisher = {IEEE Computer Society},
 address = {Washington, DC, USA},
 keywords = {fault tolerance, energy efficiency},
} 

@inproceedings{Mills_2013,
 author = {B. Mills, R. E. Grant, K. B. Ferreira and R. Riesen},
 title = {Evaluating Energy Savings for Checkpoint/Restart},
 booktitle = {Proceedings of the 1st International Workshop on Energy Efficient Supercomputing},
 series = {E2SC '13},
 year = {2013},
 isbn = {978-1-4503-2504-2},
 location = {Denver, Colorado},
 pages = {6:1--6:8},
 articleno = {6},
 numpages = {8},
 acmid = {2536432},
 publisher = {ACM},
 address = {New York, NY, USA},
 keywords = {checkpointing, energy, energy saving, fault tolerance, power, power saving},
} 
@inproceedings{DMTCP,
  title={DMTCP: Transparent checkpointing for cluster computations and the desktop},
  author={J. Ansel, K. Arya and G. Cooperman},
  booktitle={2009 IEEE International Symposium on Parallel \& Distributed Processing},
  pages={1--12},
  year={2009},
  organization={IEEE}
}

@INPROCEEDINGS{Diouri, 
author={M. Diouri, O. Gluck, L. Lefevre and F. Cappello}, 
booktitle={Dependable Systems and Networks Workshops (DSN-W), 2012 IEEE/IFIP 42nd International Conference on}, 
title={Energy considerations in checkpointing and fault tolerance protocols}, 
year={2012}, 
month={June}, 
pages={1-6}, 
keywords={checkpointing;fault tolerance;input-output programs;mainframes;power consumption;protocols;random-access storage;system monitoring;HDD logging;I-O operations;RAM logging;checkpointing;data volumes;energy consumption;energy point of view;exascale supercomputers;fault tolerance protocols;main atomic operations;power consumption;process coordination;Checkpointing;Energy consumption;Fault tolerance;Fault tolerant systems;Power demand;Protocols;Random access memory;Checkpointing;Energy consumption;Evaluation;Fault tolerance protocols}
}
@article{Bergman08exascalecomputing,
  title={Exascale computing study: Technology challenges in achieving exascale systems},
  author={K. Bergman, S. Borkar, D. Campbell, W. Carlson, W. Dally, et al.},
  journal={Defense Advanced Research Projects Agency Information Processing Techniques Office (DARPA IPTO), Tech. Rep},
  volume={15},
  year={2008}
}

@article{sankaran2005lam,
  title={The LAM/MPI checkpoint/restart framework: System-initiated checkpointing},
  author={S. Sankaran, J. M. Squyres, B. Barrett, V. Sahay, A. Lumsdaine, J. Duell, P. Hargrove and E. Roman},
  journal={The International Journal of High Performance Computing Applications},
  volume={19},
  number={4},
  pages={479--493},
  year={2005},
  publisher={Sage Publications Sage CA: Thousand Oaks, CA}
}
@article{Elnozahy2002,
  title={A survey of rollback-recovery protocols in message-passing systems},
  author={E. N. Elnozahy, L. Alvisi, Y. Wang and D. B. Johnson},
  journal={ACM Computing Surveys (CSUR)},
  volume={34},
  number={3},
  pages={375--408},
  year={2002},
  publisher={ACM}
}
@article{Elnozahy2002,
  title={A survey of rollback-recovery protocols in message-passing systems},
  author={E. N. Elnozahy, L. Alvisi, Y. Wang and D. B. Johnson},
  journal={ACM Computing Surveys (CSUR)},
  volume={34},
  number={3},
  pages={375--408},
  year={2002},
  publisher={ACM}
}

@inproceedings{mills2014shadow,
  title={Shadow Computing: An energy-aware fault tolerant computing model.},
  author={B. Mills, T. Znati and R. G. Melhem},
  booktitle={ICNC},
  pages={73--77},
  year={2014}
}
@inproceedings{LeSueur2010,
  title={Dynamic voltage and frequency scaling: The laws of diminishing returns},
  author={E. Le Sueur and G. Heiser},
  booktitle={Proceedings of the 2010 international conference on Power aware computing and systems},
  pages={1--8},
  year={2010}
}
@inproceedings{le2010dynamic,
  title={Dynamic voltage and frequency scaling: The laws of diminishing returns},
  author={Le Sueur, Etienne and Heiser, Gernot},
  booktitle={Proceedings of the 2010 international conference on Power aware computing and systems},
  pages={1--8},
  year={2010}
}
@article{Moran2018,
    author = {M. Morán, J. Balladini, D. Rexachs, E. Luque},
    title = {Factores que afectan el consumo energético de operaciones de checkpoint y restart en clusters},
    journal = {Congreso Argentino de Ciencias de la Computación},
    year = {2018},
}

@inproceedings{Silveira2016,
 author = {Silveira, D. S., Moro, G. B., Cruz, E. H. M., Navaux, P. O. A., Schnorr, L. M., and Bampi, S.},
 title = {Energy Consumption Estimation in Parallel Applications: an Analysis in Real and Theoretical Models},
 booktitle = {XVII Simposio em Sistemas Computacionais de Alto Desempenho},
 year = {2016},
 pages={134–145},
 publisher = {WSCAD},

}
@MISC{acpi,
	title = {ACPI - Advanced Configuration and Power Interface},
	howpublished = "Available at: \url{http://www.acpi.info}",
	note= {Accessed on 2018-07-09}
}

@inproceedings{Moro2018,
 author = {Gabriel B. Moro and Lucas Mello Schnorr},
 title = {Análise de Características Comportamentais de Aplicações OpenMP para Redução do Consumo de Energia},
 booktitle = {17º Workshop em Desempenho de Sistemas Computacionais e de Comunicação (WPerformance 2018)},

 volume = {17},
 location = {1/2018},
 year = {2018},
 keywords = {},
 issn = {2595-6167},
 publisher = {SBC},
 address = {Porto Alegre, RS, Brasil},
 url = {http://portaldeconteudo.sbc.org.br/index.php/wperformance/article/view/3346}
}


\begin{thebibliography}{6}
%
\bibitem {Amrizal2017}
Amrizal, M. A., Takizawa, H.: Optimizing Energy Consumption on HPC Systems with a Multi-Level Checkpointing Mechanism. In International Conference on Networking, Architecture, and Storage (NAS). IEEE (2017).

\bibitem {Mills2013}
Mills, B., Grant, R. E., Ferreira, K. B.: Evaluating energy savings for checkpoint/restart. In Proceedings of the 1st International Workshop on Energy Efficient Supercomputing. ACM (2013)

\bibitem {Mills2014}
Mills, B., Znati, T., Melhem, R., Ferreira, K. B., Grant, R. E.: Energy consumption of resilience mechanisms in large scale systems. In: 122nd Euromicro International Conference on Parallel, Distributed, and Network-Based Processing. IEEE, (2014)

\bibitem {Meneses}
Meneses, E., Sarood, O., Kalé, L. V.: Assessing energy efficiency of fault tolerance protocols for HPC systems. In 2012 IEEE 24th International Symposium on Computer Architecture and High Performance Computing (pp. 35-42). IEEE.

\bibitem {Diouri}
Diouri, M., Glück, O., Lefevre, L., Cappello, F.: Energy considerations in checkpointing and fault tolerance protocols. In IEEE/IFIP International Conference on Dependable Systems and Networks Workshops (DSN 2012) (pp. 1-6). IEEE. (2012, June).

\bibitem {Shalf2010}
Bergman, K., Borkar, S., Campbell, D., Carlson, W., Dally, W., Denneau, M., Karp, S.:  Exascale computing study: Technology challenges in achieving exascale systems. Defense Advanced Research Projects Agency Information Processing Techniques Office (DARPA IPTO), Tech. Rep, (2008)

\bibitem {Dauwe2017}
Dauwe, D., Jhaveri, R., Pasricha, S., Maciejewski, A. A., Siegel, H. J.: Optimizing checkpoint intervals for reduced energy use in exascale systems. In 2017 Eighth International Green and Sustainable Computing Conference (IGSC) (pp. 1-8). IEEE.(2017, October)

\bibitem {Rajachandrasekar2015}
Chandrasekar, R. R., Venkatesh, A., Hamidouche, K., Panda, D. K.: Power-check: An energy-efficient checkpointing framework for HPC clusters. In 2015 15th IEEE/ACM International Symposium on Cluster, Cloud and Grid Computing (pp. 261-270). IEEE. (2015, May)

\bibitem {Cui2016}
Cui, X., Znati, T., Melhem, R.: Adaptive and power-aware resilience for extreme-scale computing. In Intl IEEE Conferences on Ubiquitous Intelligence \& Computing, Advanced and Trusted Computing, Scalable Computing and Communications, Cloud and Big Data Computing, Internet of People, and Smart World Congress (UIC/ATC/ScalCom/CBDCom/IoP/SmartWorld) (pp. 671-679). IEEE.(2016, July)

\bibitem {Saito2013}
Saito, T., Sato, K., Sato, H., Matsuoka, S.: Energy-aware I/O optimization for checkpoint and restart on a NAND flash memory system. In Proceedings of the 3rd Workshop on Fault-tolerance for HPC at extreme scale (pp. 41-48). ACM. (2013, June).

\bibitem {ibtesham2014}
Ferreira, K. B., Ibtesham, D., DeBonis, D., Arnold, D.: Coarse-grained Energy Modeling of Rollback/Recovery Mechanisms (No. SAND2014-2159C). Sandia National Lab.(SNL-NM), Albuquerque, NM (United States) (2014).

\bibitem {Silveira2016}
Silveira, D. S., Moro, G. B., Cruz, E. H. M., Navaux, P. O. A., Schnorr, L. M., and Bampi, S.:Energy Consumption Estimation in Parallel Applications: an Analysis in Real and Theoretical Models. In XVII Simposio em Sistemas Computacionais de Alto Desempenho, (pp. 134-145) (2016).

\bibitem {le2010dynamic}
Le Sueur, E., Heiser, G.: Dynamic voltage and frequency scaling: The laws of diminishing returns. In Proceedings of the 2010 international conference on Power aware computing and systems (pp. 1-8).(2010, October).

\bibitem {DMTCP}
Ansel, J., Arya, K., Cooperman, G.: DMTCP: Transparent checkpointing for cluster computations and the desktop. In 2009 IEEE International Symposium on Parallel \& Distributed Processing (pp. 1-12). IEEE (2009, May).

\bibitem {Moran2018}
Morán, M., Balladini, J., Rexachs, D., Luque E.: Factores que afectan el consumo energético de operaciones de checkpoint y restart en clusters. In: XIX Workshop Procesamiento Distribuido y
Paralelo (WPDP), XXIV Congreso Argentino de Ciencias de la Computación, pp. 63-72, CACIC 2018. ISBN 978-950-658-472-6.

\bibitem {Diouri2013}
Diouri, M., Glück, O., Lefevre, L., Cappello, F.: ECOFIT: A framework to estimate energy consumption of fault tolerance protocols for HPC applications. In 13th IEEE/ACM International Symposium on Cluster, Cloud, and Grid Computing (pp. 522-529). IEEE (2013, May).

\bibitem{Bergman08exascalecomputing}
Bergman, K., Borkar, S., Campbell, D., Carlson, W., Dally, W., Denneau, M., Karp, S.: Exascale computing study: Technology challenges in achieving exascale systems. Defense Advanced Research Projects Agency Information Processing Techniques Office (DARPA IPTO), Tech. Rep, 15. (2008).

\bibitem {acpi}
ACPI - Advanced Configuration and Power Interface. \url{http://www.acpi.info}


\end{thebibliography}

\end{document}